\documentclass[runningheads,a4paper]{llncs}

\title{Gene expression modelling across multiple cell-lines with MapReduce}
\author{David M. Budden$^{1,2}$ and Edmund J. Crampin$^{1,3,4,5}$}

\institute{$^1$ Systems Biology Laboratory, Melbourne School of Engineering, The University of Melbourne, VIC 3010, Australia\\
$^2$ National ICT Australia (NICTA) Victorian Research Laboratory, Parkville, VIC 3010, Australia\\
$^3$ ARC Centre of Excellence in Convergent Bio-Nano Science and Technology, The University of Melbourne, VIC 3010, Australia\\
$^4$ Department of Mathematics and Statistics, The University of Melbourne, VIC 3010, Australia\\
$^5$ School of Medicine, The University of Melbourne, VIC 3010, Australia\\
        \email{david.budden@unimelb.edu.au}\\}

\date{}

\usepackage[table]{xcolor}
\usepackage{times}
\usepackage{url}

\usepackage{graphicx}
\usepackage{amsmath}

\usepackage{epstopdf}
\usepackage{algorithm}
\usepackage[noend]{algpseudocode}
\usepackage[makeroom]{cancel}

\begin{document}
\maketitle

\begin{abstract}
With the wealth of high-throughput sequencing data generated by recent large-scale consortia, predictive gene expression modelling has become an important tool for integrative analysis of transcriptomic and epigenetic data. However, sequencing data-sets are characteristically large, and previously modelling frameworks are typically inefficient and unable to leverage multi-core or distributed processing architectures. In this study, we detail an efficient and parallelised MapReduce implementation of gene expression modelling. We leverage the computational efficiency of this framework to provide an integrative analysis of over fifty histone modification data-sets across a variety of cancerous and non-cancerous cell-lines. Our results demonstrate that the genome-wide relationships between histone modifications and mRNA transcription are lineage, tissue and karyotype-invariant, and that models trained on matched epigenetic/transcriptomic data from non-cancerous cell-lines are able to predict cancerous expression with equivalent genome-wide fidelity.
\end{abstract}

\section{Introduction}

Computational frameworks for modelling gene expression as a function of gene-localised epigenetic features are becoming increasingly common in life sciences research. Previous studies by our lab~\cite{budden2014predictive,budden2014predicting,budden2015modelling} and others~\cite{karlic2010histone,ouyang2009chip} have leveraged the statistical power of modelling genes as observations of regulatory activity (versus variables in network-based analyses~\cite{hurley2014nail}) to gain new insight into the function and interactions of transcription factors, histone modifications and DNA methylation. Recent applications include: inference of transcription factor roles from their respective binding motifs~\cite{mcleay2012genome}; identification of regulatory elements responsible for differential expression patterns~\cite{cheng2011modeling}; exploring the relationship between gene expression and chromatin organisation~\cite{budden2014predicting}; and comparative analysis of the transcriptome across distant species~\cite{gerstein2014comparative}.

Despite the wealth of high-throughput sequencing data made available by recent large-scale consortia (\emph{e.g.} ENCODE), previous predictive modelling studies have focused on a very small number of cell-lines (typically 1-to-3~\cite{mcleay2012genome,cheng2011modeling}) despite the obvious benefits of broader, integrative analyses. We attribute this to the size of sequencing data, the computational complexity of na\"{\i}ve model implementations and widespread inability of published frameworks to leverage multi-core and/or distributed architectures. In this study, we apply the MapReduce programming paradigm~\cite{dean2008mapreduce} with domain-specific complexity optimisations to provide efficient, parallelised gene expression modelling.

A recent study by Jiang \emph{et al.} has suggested that RNA- (transcriptomic) and ChIP-seq (epigenetic) data generated in the same condition (\emph{i.e.} the same cell-line) introduces statistical bias and that specialised methods are necessary for accurately modeling the expression of cancer cells~\cite{jiang2015inference}. This study investigates both of these concerns, exploiting the computational efficiency of our MapReduce implementation and conducting an integrative analysis of six histone modifications across eight dissimilar ENCODE cell-lines. Firstly, we extend our predictive modelling framework to include $L^2$-regularisation, which is specifically designed to prevent over-fitting to experimental noise rather than meaningful biological relationships. We then quantify the extent of condition-specific bias by training and testing models on all 64 directed, pairwise combinations of cell-lines.

\section{Materials and Methods}

\subsection{ENCODE cell-line data}

Matched mRNA transcript abundance (RNA-seq) and histone modification (ChIP-seq) data were downloaded from ENCODE~\cite{encode2012integrated} for the eight cell-lines summarised in Table~\ref{tab:celllines}. These dissimilar cell-lines are the full set for which data is available for the histone modifications listed in Table~\ref{tab:hms}. The remaining histone modifications available from ENCODE are unsuitable for this study as they assert their functional role in non-promoter regions (\emph{e.g.} H3K36me3 in the $3^\prime$-UTR). The MapReduce implementation of gene expression modelling presented in this study could be trivially extended to model more cell-lines if the data were made available.

\begin{table}[]
\centering
\caption{\label{tab:celllines}All ENCODE cell-lines for which matched ChIP-seq data was available for the full set of histone modifications considered in this study (listed in Table~\ref{tab:hms}).}
\begin{tabular}{l|c|l|l|l|l}
{\bf Cell-line} & \multicolumn{1}{l|}{{\bf Tier}} & {\bf Description}                & {\bf Lineage}   & {\bf Tissue}        & {\bf Karyotype} \\ \hline
{\bf A549}      & 2                               & Alveolar carcinoma               & Endoderm        & Epithelium          & Cancer          \\
{\bf GM12878}   & 1                               & B-lymphocyte                     & Mesoderm        & Blood               & Normal          \\
{\bf H1-hESC}   & 1                               & Embryonic stem cells             & Inner cell mass & Embryonic stem cell & Normal          \\
{\bf HeLa-S3}   & 2                               & Cervical carcinoma               & Ectoderm        & Cervix              & Cancer          \\
{\bf HepG2}     & 2                               & Hepatocellular carcinoma         & Endoderm        & Liver               & Cancer          \\
{\bf HUVEC}     & 2                               & Umbilical vein endothelial cells & Mesoderm        & Blood vessel        & Normal          \\
{\bf K562}      & 1                               & Leukemia                         & Mesoderm        & Blood               & Cancer          \\
{\bf NHEK}      & 3                               & Epidermal keratinocytes          & Ectoderm        & Skin                & Normal
\end{tabular}
\end{table}

\begin{table}[]
\centering
\caption{\label{tab:hms}All histone modifications considered in this study. The remaining histone modifications available from ENCODE are unsuitable for this study as they assert their functional role in non-promoter regions (\emph{e.g.} H3K36me3 in the $3^\prime$-UTR).}
\label{my-label}
\begin{tabular}{l|l|l}
{\bf Histone modification} & {\bf Regulatory role} & {\bf Chromatin localisation}           \\ \hline
{\bf H2A.Z}                & Bivalency             & Euchromatin                  \\
{\bf H3K4me3}              & Activator/Bivalency   & Euchromatin                  \\
{\bf H3K9ac}               & Activator             & Euchromatin                  \\
{\bf H3K9me3}              & Repressor             & Constitutive heterochromatin \\
{\bf H3K27ac}              & Activator             & Euchromatin                  \\
{\bf H3K27me3}             & Repressor/Bivalency   & Facultative heterochromatin
\end{tabular}
\end{table}

\subsection{MapReduce}

MapReduce is programming paradigm which adapts the map-reduce functional programming construct for distributed and fault-tolerant data processing on commodity hardware. First developed by Google~\cite{dean2008mapreduce}, MapReduce is now widely used by companies such as Amazon, Facebook, Google and Yahoo! for parallelised processing of data on terabyte and petabyte scales. Although many of the advanced features of MapReduce are less relevant in the life sciences domain~\cite{taylor2010overview} (\emph{e.g.} optimisation of network communications, as a single sequencing data-set can often fit in main memory), the ability to seamlessly interleave sequential and parallel processing can be leveraged to reduce the processing time of gene expression modelling linearly in the number of available CPU cores.

A program implemented using the MapReduce paradigm consists of a sequence, $\langle \mu_1, \rho_1, \mu_2, \rho_2, \dots, \mu_R, \rho_R\rangle$, of mappers ($\mu_r$) and reducers ($\rho_r$) operating over $\langle \mathrm{key; value}\rangle$ pairs. Formally, a MapReduce program executes the following steps on input $U_0$ until the final reducer ($\rho_R$) halts~\cite{karloff2010model}:

\begin{algorithm}
\caption{MapReduce$\left(\langle \mu_1, \rho_1, \mu_2, \rho_2, \dots, \mu_R, \rho_R\rangle, U_0\right)$}
\begin{algorithmic}
\For {$r = 1,2,\dots,R$}
    \State ${U}^\prime_r \gets$ \textsc{Map}($U_{r-1}$)
    \State $V_r \gets$ \textsc{Shuffle}(${U}^\prime_r$)
    %\For{$V_{k,r} \in V_r$}
        \State $U_r \gets$ \textsc{Reduce}($V_r$)
    %\EndFor
\EndFor
\Return{$U_R$}
\end{algorithmic}
\begin{algorithmic}
\Function{Map}{$U_{r-1}$}
    \State ${U}^\prime_r \gets \emptyset$
    \For{$\langle k; v\rangle \in U_{r-1}$}
    \State ${U}^\prime_r \gets {U}^\prime_r \cup \mu_r\left(\langle k; v\rangle\right)$
    \EndFor
    \Return{${U}^\prime_r$}
\EndFunction
\end{algorithmic}
\begin{algorithmic}
\Function{Shuffle}{${U}^\prime_r$}
    \State $V_r \gets \emptyset$
    \For{each unique key $k \in {U}^\prime_r$}
    \State $V_{k,r} \gets \langle k; \left\{v_1, v_2, \dots, v_M\right\} \rangle : \langle k, v_m \rangle \in {U}^\prime_r$
    \State $V_r \gets V_r \cup V_{k,r}$
    \EndFor
    \Return{$V_r$}
\EndFunction
\end{algorithmic}
\begin{algorithmic}
\Function{Reduce}{$V_r$}
    \State $U_r \gets \emptyset$
    \For{each $V_{k,r} \in V_r$}
        \State $U_r \gets U_r \cup \rho_r\left(\langle k; V_{k,r} \rangle\right)$
    \EndFor
    \Return{$U_r$}
\EndFunction
\end{algorithmic}
\end{algorithm}

\noindent{The computational benefit of MapReduce follows from its inherent parallelisability, as many instances of $\mu_r$ are able to process their $\langle \mathrm{key; value}\rangle$ simultaneously (likewise with $\rho_r$, although all instances of $\mu_{r-1}$ must halt before any $\rho_r$ can commence). The following sections detail mapper and reducer implementations for various stages of the predictive gene expression modelling pipeline described in our previous studies~\cite{budden2014predictive,budden2014predicting,budden2015modelling}.}

\subsection{Quantifying transcriptional regulatory interactions}

The strength of association between a gene, $m \in (1,2,\dots, M)$, and epigenetic feature, $n \in (1,2,\dots,N)$, can be calculated from a ChIP-seq data-set:

\begin{equation*}
x_{m,n} = \sum_{\substack{r \in R_n \\|d(r,m)| \leq d^*}} \phi\left(r, m\right),
\end{equation*}

\noindent{where $R_n$ is the set of ChIP-seq reads for $n$, $d(r,m)$ is the distance (bp) separating read $r$ from the TSS of $m$, and $\phi$ maps a gene-read pair to their strength of association. The maximum bin-width, $d^*$, is typically set to 2000 to approximate the average width of ChIP-seq binding regions. Different implementations of $\phi$ are used for histone modifications (constrained sum-of-tags) versus transcription factors (exponentially decaying affinity) due to their dissimilar ChIP-seq binding profiles~\cite{budden2014predicting}:}

\begin{equation*}
\phi(r,m) =
\begin{cases}
1 & \text{for histone modifications}\\[0.8em]
\exp{\left(-\frac{d(r,m)}{d_0}\right)} & \text{for transcription factors}
\end{cases}
\end{equation*}

\noindent{where hyperparameter $d_0$ controls the strength of exponential decay for quantifying transcription factor interactions and is typically set to $d_0 = 5000$. The resultant matrix of gene-level epigenetic scores, $\mathbf{X} \in \mathcal{R}^{M\times N}$, is then log (or arsinh)-transformed and quantile-normalised for use in a regression model.}

\noindent{Given ChIP-seq data for epigenetic feature $n$ represented in UCSC wiggle (.WIG) format:}\\

\noindent{\texttt{\textbf{variableStep}\ \ \textbf{chrom}=\emph{chrN}\ \ [\textbf{span}=\emph{windowSize}]}\\
\texttt{\emph{chromStartA\ \ \ dataValueA}}\\
\texttt{\emph{chromStartB\ \ \ dataValueB}}\\
\texttt{\emph{... etc ...\ \ \ ... etc ...}}}\\

\noindent{each column $X_{\star, n} \in \mathcal{R}^{M} : X_{\star, n} = \mathrm{col}_n\left(\mathbf{X}\right)$ of the epigenetic score matrix can be efficiently calculated using MapReduce, as demonstrated below. Similar formulations can be derived for other ChIP-seq file formats.}

\begin{algorithm}
\caption{MREpigeneticScores$(X_{\star, n})$}
\begin{algorithmic}
\Procedure{Mapper $\mu\left(\left\langle X_{\star, n}; \langle locus; value \rangle \right\rangle\right)$}{}
    \For {each gene $m$}
        \If {$|d(locus, m)| \leq d^*$}
        \State \textsc{Emit}$\left(\left\langle x_{m,n}; value\times\phi(locus, m) \right\rangle\right)$
        \EndIf
    \EndFor
\EndProcedure
\Procedure{Reducer $\rho\left(\left\langle x_{m,n}; \left\{v_1, v_2, \dots, v_K  \right\} \right\rangle\right)$}{}
    \State \textsc{Emit}$\left(\left\langle x_{m,n}; \sum_{i=1}^K v_i \right\rangle\right)$
\EndProcedure
\end{algorithmic}
\end{algorithm}

\subsection{Linear regression with least squares fitting}

Suppose $\mathbf{X} \in \mathcal{R}^{M\times N}$ is a matrix of gene-level epigenetic scores, where $M$ is the number of genes and $N$ is the number of epigenetic variables ($M \gg N$). It is commonplace to model the relationship between $\mathbf{X}$ and a vector of gene-expression values, $Y \in \mathcal{R}^{M}$, in the following form:

\begin{equation*}
\label{eq:regression}
Y = \mathbf{X}\beta + \varepsilon,
\end{equation*}

\noindent{where $\beta$ parameterises the linear relationship between gene expression and epigenetics, and $\varepsilon$ are the gene-specific errors. Such models can be fitted using ordinary least squares:}

%\begin{align*}
%\hat{\beta} &= \operatornamewithlimits{argmin}_{\beta \in \mathcal{R}^{N}}\left(\left(\mathbf{X}\beta - Y\right)^\top\left(\mathbf{X}\beta - Y\right)\right)\\
%&= \left(\mathbf{X}^\top\mathbf{X}\right)^{-1}\mathbf{X}^\top Y,
%\end{align*}

\begin{align*}
\hat{\beta} &= \operatornamewithlimits{argmin}_{\beta \in \mathcal{R}^{N}} \left( || Y - \mathbf{X}\beta ||^2 \right)\\
&= \left(\mathbf{X}^\top\mathbf{X}\right)^{-1}\mathbf{X}^\top Y,
\end{align*}

\noindent{yielding the following model-based predictions of gene expression, $\hat{Y}$}:

\begin{equation*}
\label{eq:prediction}
\hat{Y} = \mathbf{X}\hat{\beta}.
\end{equation*}

Given two matrices, $\mathbf{A} \in\mathcal{R}^{X\times Y}$ and $\mathbf{B} \in\mathcal{R}^{Y\times Z}$, the product $\mathbf{C} \in \mathcal{R}^{X\times Z} : \mathbf{C} = \mathbf{A}\times\mathbf{B}$ can be formulated as follows:

\begin{equation*}
c_{i,k} \in \textbf{C} : c_{i,k} = A_{i,\star}^{\top}\times B_{\star, k}
\end{equation*}

\noindent{where:}

\begin{align*}
A_{i,\star} &\in \mathcal{R}^{X} : A_{i,\star} = \mathrm{col}_i\left(\mathbf{A}^\top\right),\\
B_{\star,k} &\in \mathcal{R}^{Z} : B_{\star, k} = \mathrm{col}_k\left(\mathbf{B}\right).
\end{align*}

\noindent{This formulation of matrix multiplication can be implemented using the MapReduce paradigm, as demonstrated below.}

%\begin{algorithm}
%\caption{MRMultiply$(\mathbf{A}, \mathbf{B})$}
%\begin{algorithmic}
%\Procedure{Mapper $\mu\left(\left\langle i; \left\{A_{\star,i}, B_{i,\star}\right\} \right\rangle\right)$}{}
%    \State $C_i \gets A_{\star,i} \times B_{i,\star}^\top$
%    \State \textsc{Emit}$\left( \left\langle \bullet ; C_i \right\rangle   \right)$
%\EndProcedure
%\Procedure{Reducer $\rho\left(\left\langle \bullet; \left\{ C_1, C_2, \dots, C_Y  \right\} \right\rangle\right)$}{}
%    \State $\mathbf{C} \gets \mathbf{0}_{X\times Z}$
%    \For{i = 1, 2, \dots, Y}
%        \State $\mathbf{C} \gets \mathbf{C} + C_i$
%    \EndFor
%    \State \textsc{Emit}$\left(\left\langle \bullet; \mathbf{C} \right\rangle\right)$
%\EndProcedure
%\end{algorithmic}
%\end{algorithm}

\begin{algorithm}
\caption{MRMultiply$(\mathbf{A}, \mathbf{B})$}
\begin{algorithmic}
\Procedure{Mapper $\mu_A\left(\left\langle \textbf{A}; a_{i,j} \right\rangle\right)$}{}
    \For{$k = 1, 2, \dots, Z$}
    \State \textsc{Emit}$\left(\left\langle c_{i,k}; a_{i,j} \right\rangle\right)$
    \EndFor
\EndProcedure
\Procedure{Mapper $\mu_B\left(\left\langle \textbf{B}; b_{j,k} \right\rangle\right)$}{}
    \For{$i = 1, 2, \dots, X$}
    \State \textsc{Emit}$\left(\left\langle c_{i,k}; b_{j,k} \right\rangle\right)$
    \EndFor
\EndProcedure
\Procedure{Reducer $\rho\left(\left\langle c_{i,k}; \left\{ A_{i,\star}, B_{\star, k} \right\} \right\rangle\right)$}{}
    \State \textsc{Emit}$\left(\left\langle c_{i,k}; A_{i,\star}^{\top}\times B_{\star, k} \right\rangle\right)$
\EndProcedure
\end{algorithmic}
\end{algorithm}

\noindent{Our implementation of linear regression with least squares fitting involves decomposing $\hat{\beta}$ into the product $\mathbf{A}^{-1}B$, where $\mathbf{A} \in \mathcal{R}^{N\times N} : \mathbf{A} = \mathrm{MRMultiply}(\mathbf{X}^\top, \mathbf{X})$ and $B \in \mathcal{R}^{N} : B = \mathrm{MRMultiply}(\mathbf{X}^\top, Y)$. The product $\mathbf{A}^{-1}B$ is calculated using standard matrix multiplication due to the unnecessary overhead of MapReduce for small matrices.}

\subsection{Regularised least squares regression}

Regularisation is a common method of overcoming the issue of over-fitting regression-based models to experimental noise rather than meaningful biological relationships. Regularisation involves penalising the fitted parameters, $\beta$, by an empirically-tuned hyperparameter, $\lambda$:

\begin{equation*}
\hat{\beta} = \operatornamewithlimits{argmin}_{\beta \in \mathcal{R}^{N}} \left( || Y - \mathbf{X}\beta ||^2 + \lambda ||\beta||^2 \right).
\end{equation*}

\noindent{Presuming $|| \bullet ||$ is the $L^2$ (Euclidean) norm, our MapReduce implementation can be trivially extended to support regularisation (implementing ridge regression). Specifically, given:}

\begin{equation*}
\tilde{Y} = \begin{bmatrix}
Y\\
0
\end{bmatrix},\; \tilde{\mathbf{X}} = \begin{bmatrix}
\mathbf{X}\\
\sqrt{\lambda}I_N
\end{bmatrix},
\end{equation*}

\noindent{where $I_N$ is the $N\times N$ identity matrix, it follows that:}

\begin{align*}
\hat{\beta} &= \operatornamewithlimits{argmin}_{\beta \in \mathcal{R}^{N}} \left( || \tilde{Y} - \mathbf{\tilde{X}}\beta ||^2 \right)\\
&= \left(\tilde{\mathbf{X}}^\top\tilde{\mathbf{X}}\right)^{-1}\tilde{\mathbf{X}}^\top \tilde{Y}\\
&= \left(\mathbf{X}^\top\mathbf{X} + \lambda I_n\right)^{-1}\mathbf{X}^\top Y.
\end{align*}

\noindent{This implementation yields the same asymptotic time complexity as ordinary least squares regression. Moreover, the existence theorem for general ridge regression demonstrates that it is always possible to tune $\lambda$ (\emph{e.g.} using cross-validation) to reduce the mean square error of model predictions~\cite{chawla1988existence,hoerl1970ridge}. This is particularly important when introducing a large number of epigenetic variables into a predictive model; \emph{e.g.} a systematic analysis of the roles of dozens of transcription factors from their ChIP-seq binding profiles. In this study, $\lambda$ is assigned the largest possible value such that the mean 10-fold cross-validated error is within 1 standard error of the minimum (solved iteratively).}

Unlike the $L^2$ norm, the $L^1$ norm is often used to enforce sparsity in $\beta$ under the assumption that most variables in $\mathbf{X}$ are practically unrelated to $Y$. This is less relevant in the context of gene expression modelling due to the well-established functional importance of epigenetic regulators for which ChIP-seq data is widely available. Moreover, the $L^1$ norm is not differentiable and thus not amenable to a closed-form MapReduce solution, although the parallelisation of iterative solutions is discussed elsewhere~\cite{zinkevich2010parallelized}.

%%https://inst.eecs.berkeley.edu/~ee127a/book/login/l_ols_rls_def.html

\section{Results}

\subsection{Histone modifications are predictive of gene expression in both cancerous and normal cell-lines}

\begin{figure}[t]
\begin{center}
    \includegraphics[width=0.9\textwidth]{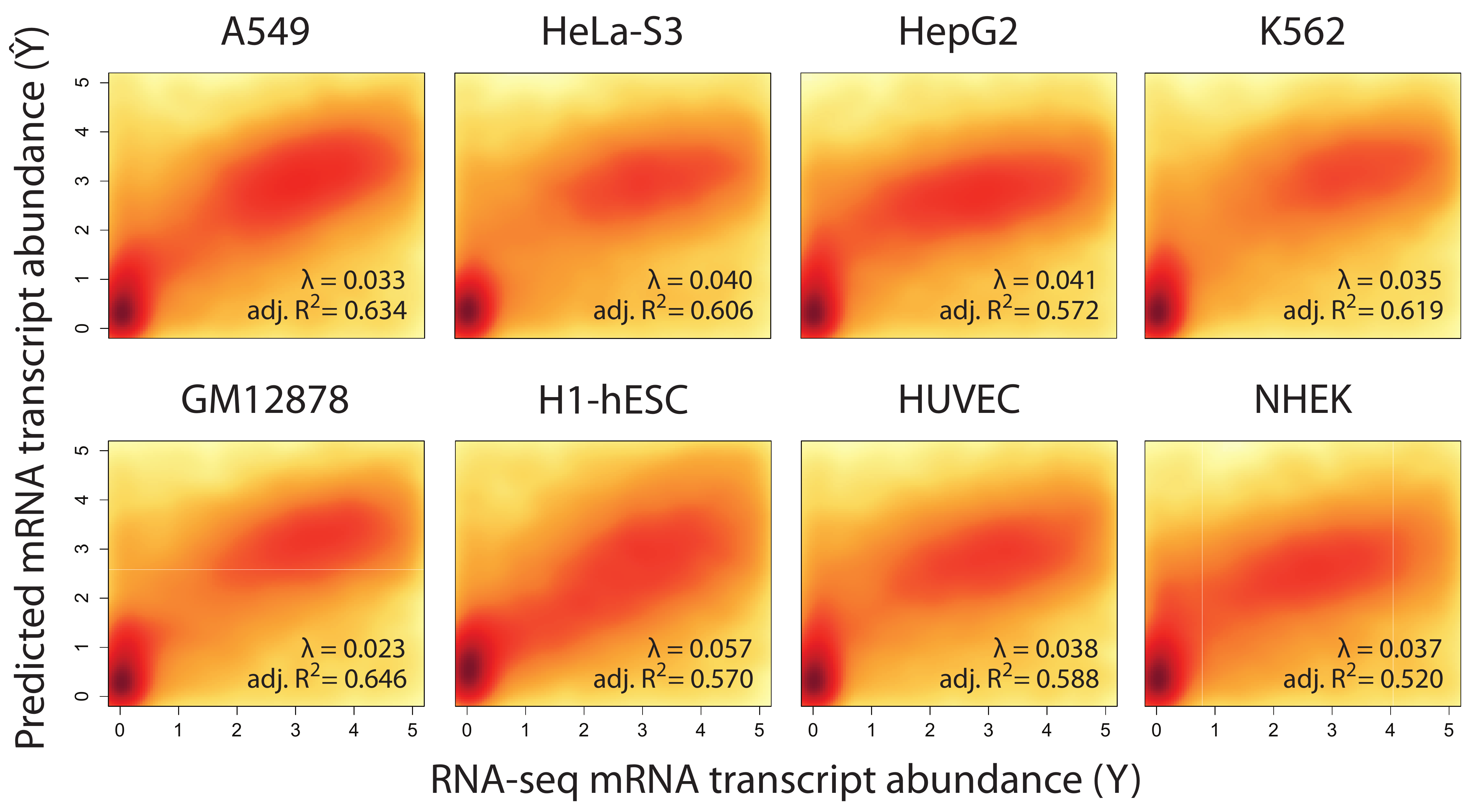}
    \end{center}
    \caption{Density plots of predicted ($\hat{Y}$) versus measured ($Y$) mRNA transcript abundance abundance for cancerous (top row, mean adj. $R^2 = 0.608$) and normal cell-lines (bottom row, mean adj. $R^2 = 0.581$). The adj. $R^2$ performance and $\lambda$ regularisation parameter (fitted using 10-fold cross validation) is reported for each cell-line.}
    \label{fig:all_scatters}
\end{figure}

$L^2$-regularised linear regression models of genome-wide mRNA transcript abundance were constructed as functions of the following histone modifications: H2A.Z, H3K4me3, H3K9ac, H3K9me3, H3K27ac and H3K27me3. For each model, the regularisation parameter, $\lambda$, was fitted using 10-fold cross-validation. The adj. $R^2$ performance of each model is presented in Fig.~\ref{fig:all_scatters}, along with a density plot of predicted ($\hat{Y}$) versus measured ($Y$) transcript abundance. It is evident that histone modifications are accurate predictors of gene expression in both cancerous (top row, mean adj. $R^2 = 0.608$) and normal cell-lines (bottom row, mean adj. $R^2 = 0.581$), despite recent studies suggesting that specialised models are necessary to appropriately model cancerous cells~\cite{jiang2015inference}.

\begin{figure}[t]
\begin{center}
    \includegraphics[width=0.9\textwidth]{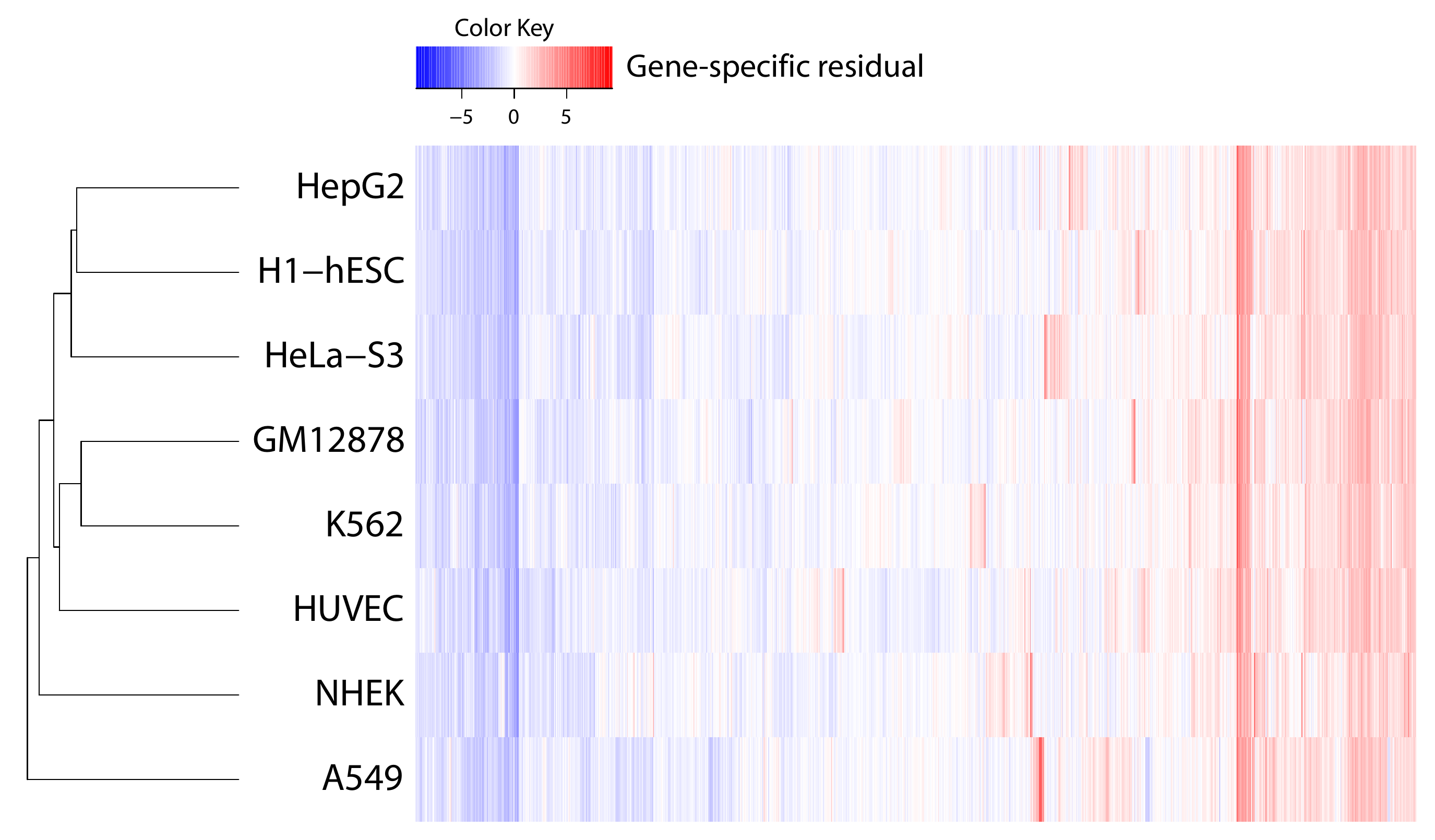}
    \end{center}
    \caption{Hierarchical clustering of cell-lines by mRNA transcript abundance residuals ($\varepsilon = Y - \hat{Y}$). The three mesodermal derivatives GM12878, K562 and HUVEC cluster together, suggesting that residuals are partially non-random and instead convey meaningful biological information. Consistently, it is evident that the expression levels of many genes are poorly predicted across all eight cell-lines, presumably capturing divergence from histone modification-mediated regulation (explored in detail in our previous study~\cite{budden2014predicting}).}
    \label{fig:line_cluster}
\end{figure}

Fig.~\ref{fig:line_cluster} presents the results of hierarchically clustering cell-lines by mRNA transcript abundance residuals ($\varepsilon = Y - \hat{Y}$). Interestingly, the three mesodermal derivatives GM12878, K562 and HUVEC form a distinct cluster. RNA-sequencing data for the least similar cell-line (A549) was generated at Cold Spring Harbor Laboratory whereas all other transcriptomic data was generated at the California Institute of Technology, suggesting that batch effects may be a contributing factor. It is also evident that the expression levels of many genes are consistently over- or under-estimated across all eight cell-lines. Taken together, these results indicate that gene-specific residuals are non-random but instead indicative of genes that are inherently difficult to model from histone modification data. The existence of genes with transcriptional activity apparently decoupled from the local epigenetic landscape has been explored in detail in our previous study~\cite{budden2014predicting}.

\subsection{The regulatory roles of histone modifications are cell-line invariant}

\begin{figure}[t]
\begin{center}
    \includegraphics[width=1.0\textwidth]{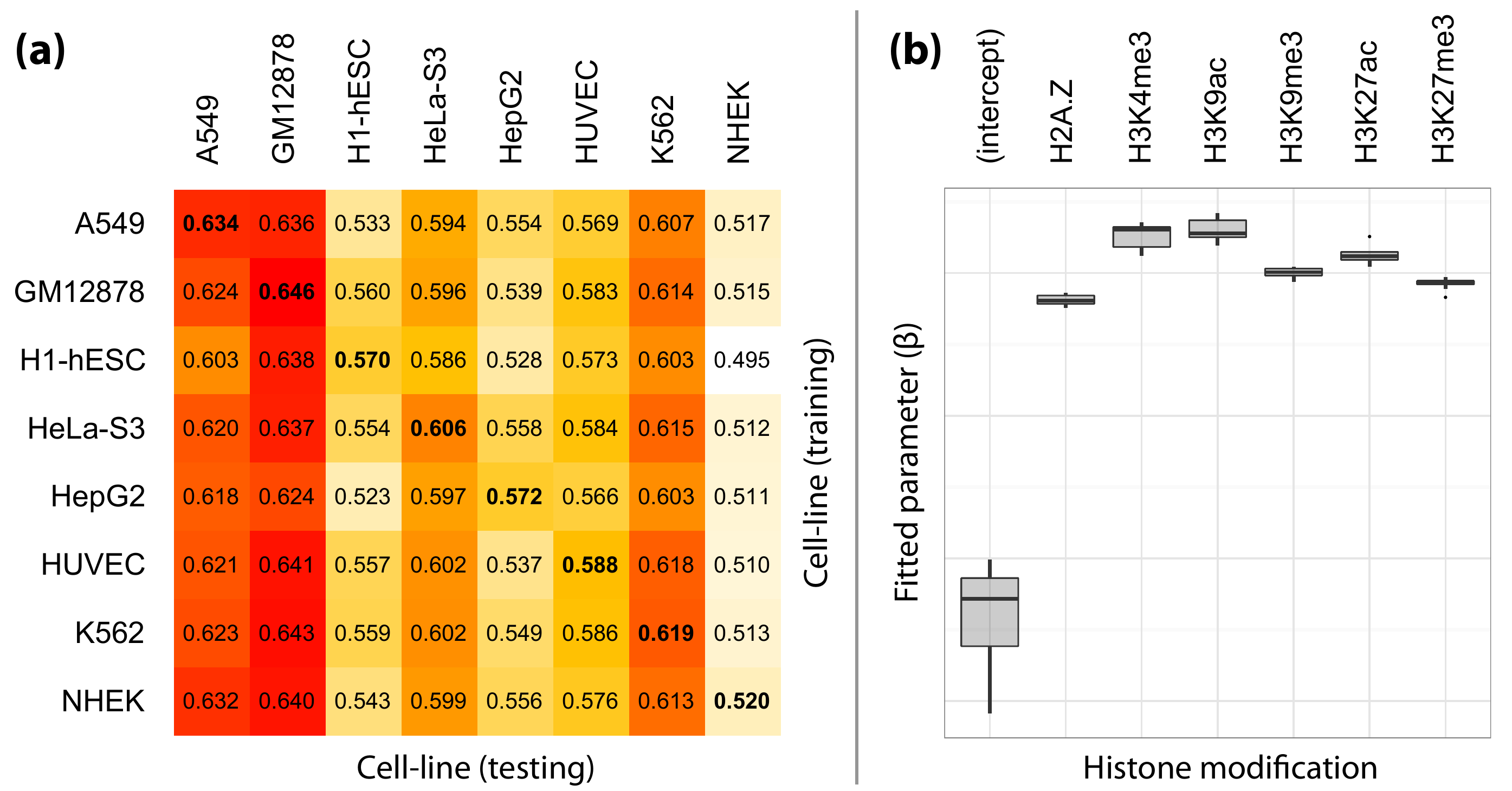}
    \end{center}
    \caption{\textbf{(a)} Genome-wide accuracy of mRNA transcript abundance predictions (adj. $R^2$) for models trained and tested on each pairwise combination of cell lines. These results are strikingly non-symmetric, with significant dissimilarity between columns (predictions) but not rows (training observations). \textbf{(b)} Distribution of each fitted model parameter, $\hat{\beta}_m$, across all cell-lines considered in this study.}
    \label{fig:pairwise_performance}
\end{figure}

To assess the extent to which condition-specific bias influences the reported accuracy of gene expression predictions, we trained and tested models on all 64 directed, pairwise combinations of cell-lines. The adj. $R^2$ performance for these models are presented in Fig.~\ref{fig:pairwise_performance}(a). These results demonstrate significant non-symmetry, with dissimilarity between columns (predictions) but not rows (training observations). This demonstrates that the transcriptional regulatory roles of histone modifications are cell line invariant at a genome-wide level (within the constraints of a linear model); \emph{e.g.} A549 and GM12878 expression can be accurately predicted by models trained on any cell-line, despite their diversity in lineage, tissue and karyotype. These results are further supported by Fig.~\ref{fig:pairwise_performance}(b), which demonstrates consistency in the fitted model parameters, $\hat{\beta}$, across all cell-lines.

It is worth noting that that models trained and tested using data from a single cell-line (boldfaced along the diagonal of Fig.~\ref{fig:pairwise_performance}(a)) only marginally outperform models trained on dissimilar cell-lines and, moreover, that these margins are significantly less than the inherent variation between columns. These findings suggest that, in the context of gene expression modelling, training and testing models on data generated under the same experimental conditions (\emph{i.e.} the same cell-line) is not a significant source of statistical bias.

\subsection{MapReduce reduces the asymptotic time complexity of gene expression modelling}

For $M$ genes and a ChIP-seq data-set containing $R$ mapped reads, the asymptotic time complexity class of generating a column $X_{\star, n}$ of $\mathbf{X}$ is $\Theta(MR)$. By first preprocessing the list of gene TSS loci (invariant between epigenetic datasets) into a balanced binary search tree and observing that the vast majority of reads are within $d^*$ bp of exactly zero-or-one gene, our MapReduce implementation of calculating $X_{\star, n}$ yields the following complexity when distributed across $P$ MapReduce nodes:

\begin{equation*}
\mathrm{MREpigeneticScores} \in \Theta\left( \frac{R\log(M)}{P}\right),
\end{equation*}

\noindent{which must be completed separately for each epigenetic feature, $n \in (1,2,\dots,N)$.}

For $\mathbf{X} \in \mathcal{R}^{M\times N}$ and $Y \in \mathcal{R}^{M}$, the asymptotic time complexity of ordinary least squares fitting $\hat{\beta} = f(\mathbf{X}, Y)$ can also be derived:
\begin{equation*}
f \in \underbrace{\xcancel{\Theta(MN)}}_{\mathbf{X}^\top} +  \underbrace{\Theta(MN^2)}_{\mathbf{A} = \mathbf{X}^\top\mathbf{X}} + \underbrace{\xcancel{\Theta(MN)}}_{B=\mathbf{X}^\top Y} + \underbrace{\Theta(N^3)}_{\mathbf{A}^{-1}} + \underbrace{\xcancel{\Theta(N^2)}}_{\mathbf{A}^{-1}B}
\end{equation*}

\noindent{Observing that $R \gg M \gg N$ for gene expression modelling and by distributing the calculation of $\mathbf{A}$ and $B$ across $P$ MapReduce nodes, the overall complexity reduces to:}

\begin{align*}
\mathrm{MRExpressionModelling} &\in \Theta\left(\frac{NR\log(M)}{P}\right)\\
\end{align*}

\noindent{thus this MapReduce implementation of gene expression modelling yields an optimal $\Theta(P)$ improvement in asymptotic time complexity without the need to parallelise matrix inversion or transpose operations.}

\section{Discussion}

Many previous predictive modelling studies have been limited in scope to 1-3 cell-lines due to the computational expense of integrating and analysing large, high-throughput sequencing data-sets. In this study, we introduced a MapReduce implementation of gene expression modelling that is able to obtain a full $\Theta(P)$ improvement in asymptotic time complexity when distributed across $P$ CPUs (\emph{e.g.} as part of multi-core PC or high-performance cluster). This implementation was subsequently applied to an integrative analysis of more than 50 epigenetic and matched transcriptomic data-sets across 8 dissimilar ENCODE cell-lines.

Despite recent studies presenting specialised methods for modelling cancerous gene expression~\cite{jiang2015inference}, we find no evidence of variation in the statistical relationship between histone modifications and mRNA transcript abundance in normal-versus-cancerous cell-lines. Although our results demonstrate that some cell-lines are inherently more difficult to model than others, this trait appears to be more closely associated with the extent of cellular differentiation than carcinogenic state; \emph{e.g.} models of h1-hESC embryonic stem cells perform 12\% worse than terminally-differentiated GM12878 lymphoblasts. Although the NHEK (Normal Human Epidermal Keratinocytes) cell-line is both terminally-differentiated and exhibits the worst-performing models, this may be attributed to the phenotypic plasticity of keratinocytes between epithelial and mesenchymal states (necessary for wound healing). We therefore speculate that the predictability of a cell-line's genome-wide expression levels from epigenetic data is proportional to its transcriptomic rigidity; \emph{i.e.} cells with signal-induced phenotypic plasticity are less likely to exhibit a stable, predictive epigenome.

%The fitted model parameters are practically invariant across all cell-lines, thus performance variation is due to variance in gene expression (presumably due to signal-induced transcriptomic plasticity) rather than cell-line specific regulatory activity at the level of histone modifications.

Interestingly, hierarchical clustering of the 8 investigated cell-lines by mRNA transcript abundance residuals (gene-level prediction errors) was able to group the closely-related, mesodermal-derivative cell-lines GM12878, K562 and HUVEC; again, carcinogenic state appeared to have little effect on the propensity of two cell-lines to cluster together. Taken together with the observation that many genes exhibited large and consistent residuals across all cell-lines, these results suggest that gene-level residuals are non-random and, moreover, that the transcriptional activity of many genes are decoupled from their local epigenetic landscape~\cite{budden2014predicting}.

\section*{Acknowledgements}

This work was supported by an Australian Postgraduate Award [DMB]; the Australian Federal and Victoria State Governments and the Australian Research Council through the ICT Centre of Excellence program, National ICT Australia (NICTA) [DMB]; and the Australian Research Council Centre of Excellence in Convergent Bio-Nano Science and Technology (project number CE140100036) [EJC]. The views expressed herein are those of the authors and are not necessarily those of NICTA or the Australian Research Council.

\bibliographystyle{splncs}
\bibliography{refs}

\begin{thebibliography}{10}

\bibitem{budden2014predictive}
Budden, D.M., Hurley, D.G., Crampin, E.J.:
\newblock Predictive modelling of gene expression from transcriptional
  regulatory elements.
\newblock Briefings in Bioinformatics (2014)  bbu034

\bibitem{budden2014predicting}
Budden, D.M., Hurley, D.G., Cursons, J., Markham, J.F., Davis, M.J., Crampin,
  E.J.:
\newblock Predicting expression: the complementary power of histone
  modification and transcription factor binding data.
\newblock Epigenetics \& Chromatin \textbf{7}(36) (2014)  1--12

\bibitem{budden2015modelling}
Budden, D.M., Hurley, D.G., Crampin, E.J.:
\newblock Modelling the conditional regulatory activity of methylated and
  bivalent promoters.
\newblock Epigenetics \& Chromatin \textbf{8}(21) (2015)

\bibitem{karlic2010histone}
Karli{\'c}, R., Chung, H.R., Lasserre, J., Vlahovi{\v{c}}ek, K., Vingron, M.:
\newblock Histone modification levels are predictive for gene expression.
\newblock Proceedings of the National Academy of Sciences \textbf{107}(7)
  (2010)  2926--2931

\bibitem{ouyang2009chip}
Ouyang, Z., Zhou, Q., Wong, W.H.:
\newblock {ChIP-Seq of transcription factors predicts absolute and differential
  gene expression in embryonic stem cells}.
\newblock Proceedings of the National Academy of Sciences \textbf{106}(51)
  (2009)  21521--21526

\bibitem{hurley2014nail}
Hurley, D.G., Cursons, J., Wang, Y.K., Budden, D.M., Crampin, E.J.,  et~al.:
\newblock {NAIL, a software toolset for inferring, analyzing and visualizing
  regulatory networks}.
\newblock Bioinformatics (2014)  btu612

\bibitem{mcleay2012genome}
McLeay, R.C., Lesluyes, T., Partida, G.C., Bailey, T.L.:
\newblock Genome-wide in silico prediction of gene expression.
\newblock Bioinformatics (2012)  bts529

\bibitem{cheng2011modeling}
Cheng, C., Gerstein, M.:
\newblock Modeling the relative relationship of transcription factor binding
  and histone modifications to gene expression levels in mouse embryonic stem
  cells.
\newblock Nucleic Acids Research (2011)  gkr752

\bibitem{gerstein2014comparative}
Gerstein, M.B., Rozowsky, J., Yan, K.K., Wang, D., Cheng, C., Brown, J.B.,
  Davis, C.A., Hillier, L., Sisu, C., Li, J.J.,  et~al.:
\newblock Comparative analysis of the transcriptome across distant species.
\newblock Nature \textbf{512}(7515) (2014)  445--448

\bibitem{dean2008mapreduce}
Dean, J., Ghemawat, S.:
\newblock Mapreduce: simplified data processing on large clusters.
\newblock Communications of the ACM \textbf{51}(1) (2008)  107--113

\bibitem{jiang2015inference}
Jiang, P., Freedman, M.L., Liu, J.S., Liu, X.S.:
\newblock Inference of transcriptional regulation in cancers.
\newblock Proceedings of the National Academy of Sciences \textbf{112}(25)
  (2015)  7731--7736

\bibitem{encode2012integrated}
{ENCODE Project Consortium},  et~al.:
\newblock {An integrated encyclopedia of DNA elements in the human genome}.
\newblock Nature \textbf{489}(7414) (2012)  57--74

\bibitem{taylor2010overview}
Taylor, R.C.:
\newblock {An overview of the Hadoop/MapReduce/HBase framework and its current
  applications in bioinformatics}.
\newblock BMC Bioinformatics \textbf{11}(Suppl 12) (2010) ~S1

\bibitem{karloff2010model}
Karloff, H., Suri, S., Vassilvitskii, S.:
\newblock {A model of computation for MapReduce}.
\newblock In: Proceedings of the Twenty-First Annual ACM-SIAM Symposium on
  Discrete Algorithms, Society for Industrial and Applied Mathematics (2010)
  938--948

\bibitem{chawla1988existence}
Chawla, J.:
\newblock The existence theorem in general ridge regression.
\newblock Statistics \& Probability Letters \textbf{7}(2) (1988)  135--137

\bibitem{hoerl1970ridge}
Hoerl, A.E., Kennard, R.W.:
\newblock Ridge regression: Biased estimation for nonorthogonal problems.
\newblock Technometrics \textbf{12}(1) (1970)  55--67

\bibitem{zinkevich2010parallelized}
Zinkevich, M., Weimer, M., Li, L., Smola, A.J.:
\newblock Parallelized stochastic gradient descent.
\newblock In: Advances in Neural Information Processing Systems, Neural
  Information Processing Systems Foundation (2010)  2595--2603

\end{thebibliography}
\end{document}